\documentclass[%
reprint,
 amsmath,amssymb,
 aps,
longbibliography 
]{revtex4-2}
\usepackage{ragged2e}
\usepackage{graphicx}
\usepackage{dcolumn}
\usepackage{bm}
\usepackage{tabularx,array} 
\newcolumntype{Y}{>{\centering\arraybackslash}X} 
\begin{document}


\title{Universal Loop Statistics from Active Extrusion with Kinetic Barriers}

\author{A. Chervinskaya$^{1}$} 
\author{R. Metzler$^{2}$} 
\author{K. E. Polovnikov$^{1,2,\;}$} \email{kipolovnikov@gmail.com} 

\affiliation{$^1$ AI Center, Skolkovo Institute of Science and Technology, 121205 Moscow, Russia \\
$^2$ Institute for
Physics \& Astronomy, University of Potsdam, 14476 Potsdam-Golm, Germany}

\date{\today}

\begin{abstract}
We develop a kinetic theory of cohesin-driven loop extrusion on a disordered chromatin track with transient barriers. In the stationary state, the mean loop size is shown to obey a universal law determined by the bare processivity and a renormalized obstacle density. Beyond the mean, one-sided extrusion always yields a single-exponential loop-length distribution, whereas two-sided
extrusion produces a finite sum of exponential modes and, generically, a
peaked distribution. 
Experimental CTCF-anchored loop statistics exhibit such a peak,
thereby providing a direct discriminator of extrusion symmetry. The theory therefore establishes a unified framework for disorder-limited loop extrusion and supports a scenario in which both cohesin arms actively operate in living cells.
\end{abstract}

\maketitle

\textit{Introduction}. 
Folding meter-long human DNA into a micrometer-sized nucleus is a long-standing
problem of modern soft condensed matter and biophysics \cite{dekker24,Fraser07}. A major contribution to this cross-scale compaction comes from chromatin loop formation
by active extrusion, in which cohesin---a chromosome-organizing motor of the
structural-maintenance-of-chromosomes (SMC) family---binds chromatin and
enlarges a loop until it dissociates or is halted by chromatin-bound obstacles, such as CTCF, a sequence-specific DNA-binding factor that acts as a major
positional barrier (``roadblock'') to extrusion \cite{sanborn15,fudenberg2016,brackley17,rao17,mach22,gabriele22,Sabate25,brandao19,dequeker22}.
From the viewpoint of statistical physics, loop extrusion is a driven one-dimensional process in a dynamically disordered environment.

One-dimensional disordered systems are paradigmatic models of nonequilibrium statistical physics. Random walks in random environments exhibit trapping, localization,
and anomalous diffusion \cite{metzler00,Goychuk14prl,Goychuk14,Fisher98,Burov07}, with Sinai diffusion as a classic example of disorder-dominated transport
in one dimension \cite{sinai83,bouchaud90,Padash22,Oshanin13}. Driven lattice gases and exclusion processes with quenched or annealed disorder likewise show how heterogeneity reshapes fluctuations and steady states \cite{derrida98,evans00,chou11}.
Loop extrusion belongs naturally to this class, but with an important twist: a cohesin motor extrudes only while it remains bound to chromatin for a characteristic time \(\tau\). Unlike Sinai diffusion, where
the typical displacement grows logarithmically in time, cohesin exchange therefore cuts off the motion and thereby sets a finite loop scale given by the intrinsic processivity \(l_p=v\tau\), where \(v\) is the extrusion speed \cite{fudenberg2016}.

\begin{figure}[ht]
\centering
\includegraphics[width=0.47\textwidth]{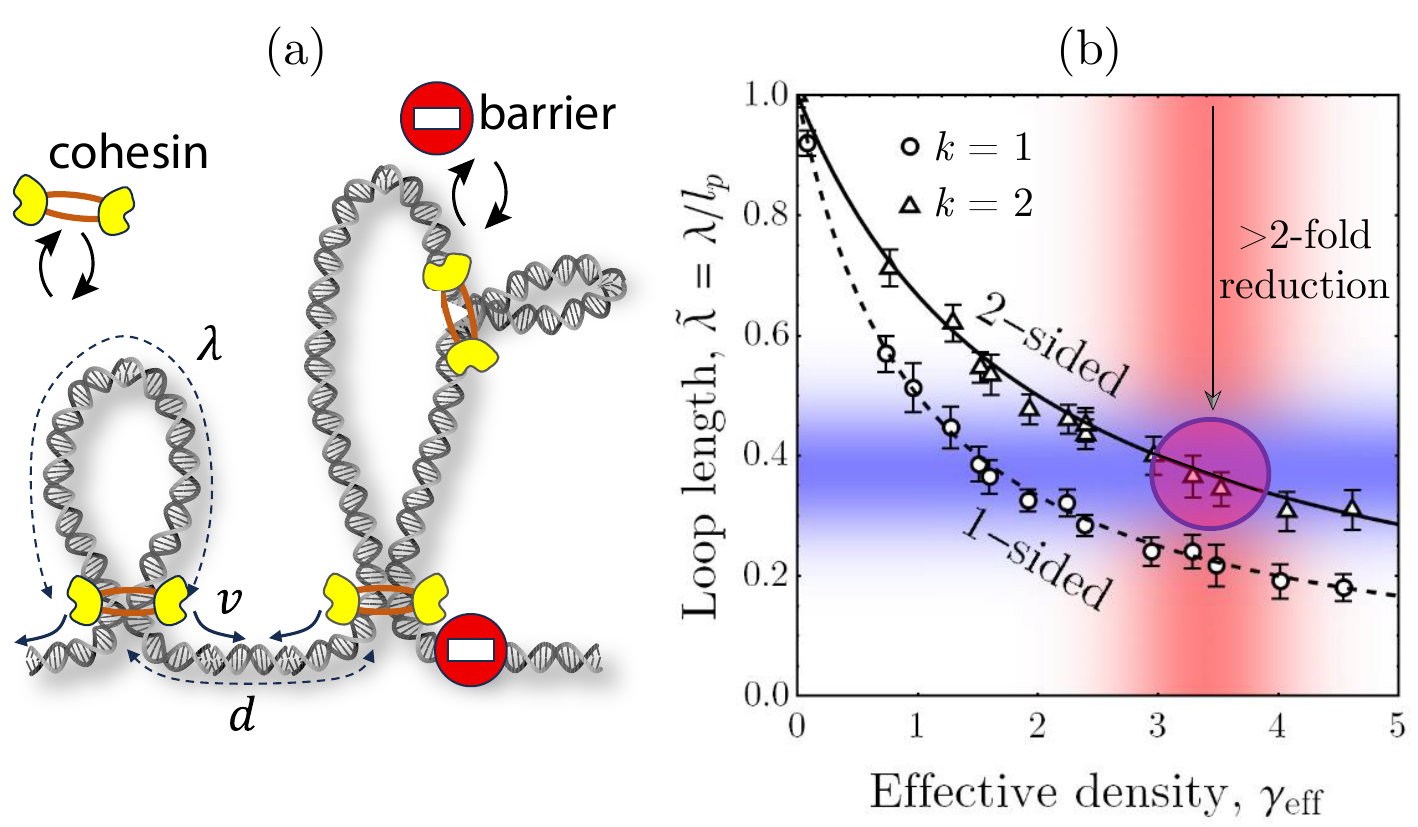}
\caption{Universal disorder-induced reduction of loop length. (a) Schematic of two-sided loop extrusion on chromatin. A cohesin motor with intrinsic lifetime $\tau$ on the chain and processivity \(l_p=v\tau\) extrudes a loop at total speed \(v\), but encounters stochastically bound roadblocks and other extruders that limit the mean loop length to \(\lambda<l_p\). The mean spacing \(d\) between cohesins characterizes extruder crowding factor \(\gamma=l_p/d\). (b) Reduced mean loop length \(\tilde{\lambda}=\lambda/l_p\) versus effective obstacle density \(\gamma_{\mathrm{eff}}\). Solid curve: two-sided extrusion, \(\tilde{\lambda}=1/(1+\gamma_{\mathrm{eff}}/2)\); dashed curve: one-sided extrusion, \(\tilde{\lambda}=1/(1+\gamma_{\mathrm{eff}})\). Symbols: stochastic simulations. The blue horizontal band shows the experimental range \(\tilde{\lambda}=\langle v\rangle/v\) for HeLa G1 \cite{rao17,davidson19,kim19}; the red vertical band shows the range of \(\gamma_{\mathrm{eff}}\) inferred from measured barrier densities and lifetimes (Table~S1). Their overlap (magenta circle) corresponds to the two-sided branch.}
\label{fig:fig1}
\end{figure}

Loop extrusion [Fig.~1(a)] was proposed as the physical mechanism underlying domain formation in Hi-C contact maps \cite{sanborn15,fudenberg2016}, later confirmed by single-molecule in vitro experiments \cite{kim19,davidson19,ganji18}, and has since become a central ingredient of chromosome-folding models \cite{nuebler18,lizana24,bonato24,forte24,Stefano22}. Most theoretical work, however, has focused on the three-dimensional consequences of extrusion, through simulations of chromosome mechanics and contact patterns \cite{nomidis22,takaki21,tortora25,salari26,banigan2020,chan23,pinto26,chan24}. At the same time, key properties of extrusion in living cells remain unresolved, most notably its symmetry---whether cohesin extrudes effectively in a one-sided
or two-sided manner \cite{Barth25,Oldenkamp22}. Recent analytically tractable models have successfully reproduced in vivo Hi-C contact statistics \cite{polovnikov-starkov26,polovnikov23prx,polovnikov23pre,slavov23,belan24}, but they do so by phenomenologically mapping loop disorder onto a two-state Markov process, which enforces exponential loop lengths. The central physical question therefore remains open: what stationary loop-length statistics are generated by active extrusion itself on a one-dimensional disordered chromatin track, and how do they depend on extrusion symmetry?

In what follows, we develop a general mean-field kinetic theory of active loop extrusion in the presence of arbitrary number of transiently bound roadblocks. We first derive a
universal expression for the mean loop size \(\lambda\), and then show that the full loop-length distribution \(N(x)\) is far more diagnostic: one-sided extrusion always yields an exponential law, whereas two-sided extrusion
generically produces a peaked distribution. Comparison with experimental
CTCF-anchored loop statistics reveals such a peak [Fig.~3(b)], thereby showing that a purely one-sided mechanism of cohesin extrusion is incompatible with the observed in vivo data.

\textit{1D Loop extrusion model}. In the standard picture \cite{fudenberg2016},
a cohesin complex loads onto chromatin, remains bound for a mean time \(\tau\),
and extrudes a loop of size \(x\) at total speed \(v\) until it dissociates or
is blocked by a chromatin-bound obstacle. We explicitly distinguish one-sided extrusion (\(k=1\), with only one cohesin arm active) from symmetric two-sided extrusion (\(k=2\)); in
both cases \(v\) denotes the total loop-growth speed.

In the absence of obstacles, the intrinsic processivity \(l_p=v\tau\) sets both the characteristic loop scale and the exponential loop-length distribution. In vivo, extrusion is hindered by \(n\) classes of transiently bound static roadblocks, \(i=1,\dots,n\), characterized by lifetime \(\tau_i\), permeability \(p_i\), and effective separation \(d_i\) [Fig.~1(a)], defined as the mean spacing between successful blocking events \footnote{If a roadblock stops cohesin with probability \(p_i\) upon encounter, this partial permeability is absorbed into
the effective spacing, which exceeds the raw genomic spacing by a factor
\(p_i^{-1}\).}. Other cohesins act as a separate obstacle class with mean
effective spacing \(d\), thereby absorbing possible mutual bypassing at the mean-field level. The stationary state is then conveniently characterized by the dimensionless parameters
\begin{equation}
\gamma_i=\frac{l_p}{d_i},
\qquad
\gamma=\frac{l_p}{d},
\qquad
\alpha_i=\frac{\tau}{\tau_i},
\label{eq:gammas}
\end{equation}
where \(\gamma_i\) and \(\gamma\) are crowding parameters, while \(\alpha_i\) controls roadblock persistence: \(\alpha_i\ll1\) corresponds to long-lived
barriers, whereas \(\alpha_i\gtrsim1\) to short-lived ones; for cohesin
obstacles, \(\alpha_c=1\) by definition. 
The key idea of this Letter is that dynamic disorder reduces the mean loop size through a single renormalized obstacle density, whereas the full loop-length distribution serves as a qualitative diagnostic of extrusion symmetry $k$.

\textit{Universal mean loop size}. In the dilute limit, \(\gamma,\gamma_i\ll1\), collisions are rare and the stationary mean loop size \(\lambda\) approaches the motor processivity \(l_p\). At finite obstacle density, repeated encounters reduce \(\lambda\). Within mean field, stationary extrusion is described by state-resolved balance equations combining advection in loop length with stochastic blocking, motor release and barrier dissociation (see SM, \cite{suppl}).
Solving these equations yields, for an arbitrary number of barrier classes
\(n\),
\begin{equation}
\lambda=\frac{l_p}{1+\gamma_{\mathrm{eff}}/k},
\label{eq:lambda_universal}
\end{equation}
where the effective density \(\gamma_{\mathrm{eff}}\) is additive in all static roadblocks and other cohesins:
\begin{equation}
\gamma_{\mathrm{eff}}
=
\underbrace{\sum_{i=1}^{n}\frac{\gamma_i}{1+\alpha_i}}_{\text{static barriers}}
+\underbrace{\frac{\gamma_c}{2}}_{\text{cohesins}},
\qquad
\gamma_c=\gamma+\gamma_{\mathrm{dyn}}^{(k)}.
\label{eq:geff}
\end{equation}
Equations~\eqref{eq:lambda_universal} and \eqref{eq:geff} are our first key
result.
The first term in Eq.~\eqref{eq:geff} collects the \(n\) static roadblock
classes, while the second accounts for encounters with other extruders.
Long-lived roadblocks (\(\alpha_i\ll1\)) contribute nearly their full density \(\gamma_i\), whereas short-lived ones are suppressed by \((1+\alpha_i)^{-1}\). Thus roadblock occupancy alone is insufficient: the relevant quantity is the probability that a roadblock persists over the lifetime of the extruder.

The cohesin contribution \(\gamma_c\) contains the static term \(\gamma\) and the dynamic correction \(\gamma_{\mathrm{dyn}}^{(k)}\), both reduced by the cohesin persistence factor \((1+\alpha_c)^{-1}=1/2\). Unlike static roadblocks, this correction originates from the relative motion of neighboring
extruders and is therefore most important in the dilute limit, where
\(\gamma_{\mathrm{dyn}}^{(k)}\approx (k/2)\gamma\) for \(\gamma,\gamma_i\to0\) \cite{suppl}.
Under experimentally relevant conditions, \(\gamma\approx2.7\) (Table~S2), it remains small, \(\gamma_{\mathrm{dyn}}\approx0.4\), compared to the static term. Most cohesins therefore behave effectively as blocked obstacles with
\(\alpha_c=1\), so extruder mobility contributes only weakly to
\(\gamma_{\mathrm{eff}}\).

Accordingly, in the obstacle-free limit, \(\gamma_{\mathrm{eff}}\to0\) and \(\lambda\to l_p\). For fixed total extrusion speed \(v\), two-sided extrusion (\(k=2\)) reduces the speed of each arm to \(v/2\), halves the encounter rate, and therefore yields a larger \(\lambda\) than one-sided extrusion (\(k=1\)).
Stochastic simulations of active extrusion with explicit loading,
dissociation, and steric interactions confirm Eq.~\eqref{eq:lambda_universal}
for both \(k=1\) and \(k=2\) [Fig.~1(b)].

Equation~\eqref{eq:lambda_universal} also admits a direct kinetic interpretation for $\tilde{\lambda}=\lambda/l_p$ that is convenient for experimental tests:
\begin{equation}
\tilde{\lambda}=\langle v\rangle/v,
\end{equation}
where \(\langle v\rangle\) is the average loop-growth rate. In HeLa cells \cite{rao17}, recovery of a \(900\pm 50\) kb Hi-C loop occurs within \(\approx40\) min after cohesin restoration; together with the single-molecule measurements of the intrinsic cohesin speed \(v\approx1\,\mathrm{kb/s}\) \cite{davidson19,kim19} this yields $\tilde{\lambda} \approx 0.38 \pm 0.02$ for the relative loop size. 
Independently, using measured barrier densities and lifetimes to estimate \(\gamma_{\mathrm{eff}}\) (Table~S1), Eq.~\eqref{eq:lambda_universal} predicts
$\tilde{\lambda}\approx0.37\pm0.07\;\text{for } k=2$ and
$\tilde{\lambda}\approx0.23\pm0.05
\;\text{for } k=1 $
[Fig.~1(b)]. Thus the theory appears to be quantitatively consistent with experiment only for two-sided extrusion, indicating that obstacle disorder reduces the in vivo mean loop size to \(\approx30\)–\(50\%\) of the intrinsic processivity \(l_p\).

\begin{figure}[t]
    \centering
\includegraphics[width=0.48\textwidth]{fig2.pdf}   
    
\caption{Length-state graph structure determines loop-size statistics.
(a) One-sided extrusion: Only the free state \(A_{0}\) propagates in loop length; once the moving arm is blocked, growth stops.
(b) Two-sided extrusion: Mixed states with one blocked and one mobile arm (\(A_b\), \(A_c\)) remain propagating, in addition to $A_0$, and introduce additional decay channels.
(c) As a consequence, one-sided extrusion yields a purely exponential loop-size distribution for all $\gamma_b$.
(d) Two-sided extrusion yields a sum of exponentials and, generically, a peaked distribution. The red dashed line shows the limiting \(\Gamma\)-like form with shape parameter \(2\).
In panels (c) and (d), the theoretical curves are shown for single-type long-lived barriers (\(\alpha=0\)) of varying density $\gamma_1 \equiv \gamma_b$ at fixed extruder density \(\gamma=2.7\). Blue histograms show the results of stochastic simulations for \(\gamma_b=1.8\), corresponding to the experimental estimate (Table~S2); blue dotted lines highlight the corresponding analytical prediction.}
\label{fig:fig2}
\end{figure}

\textit{Loop-length distribution}. The mean loop size \(\lambda\) is universal and therefore only weakly diagnostic of extrusion symmetry \(k\). The full stationary distribution \(N(x)\), in contrast, sharply distinguishes one- and two-sided extrusion and provides the main experimental discriminator of the theory.


Let \(\mathbf A(x)\) be the vector of state-resolved densities at loop size \(x\) (Fig.~2). For a single barrier class, \(\mathbf A=\{A_{0},A_b,A_c\}\) for one-sided extrusion and \(\mathbf A=\{A_{0},A_b,A_c,A_{bb},A_{bc},A_{cc}\}\) for two-sided extrusion. Here \(A_{0}\) is the free state; \(A_b\) and \(A_c\) denote states with one arm blocked by a static barrier or another cohesin, respectively, while the other remains mobile; and \(A_{bb}\), \(A_{bc}\), and \(A_{cc}\) are fully blocked states. Additional obstacle classes simply enlarge this vector by adding further blocked-state combinations.

As different transitions change the loop length by different amounts, it is instructive to rewrite the continuous-time balance equations as discrete propagation in loop length,
\begin{equation}
\mathbf A(x+\Delta x)=\mathbf L\,\mathbf A(x), \qquad
\mathbf A(0)\propto \delta(x)\,\mathbf e_{0},
\label{eq:Ptransfer}
\end{equation}
where \(\mathbf L\) is an effective transfer operator associated with an
increment \(\Delta x\) in loop length $x$. Equation~\eqref{eq:Ptransfer} maps the steady-state loop statistics onto a survival problem on the directed length-state graph \(\mathbf L\) with point-like initial condition
[Fig.~2(a,b)]: probability is injected into the root node \(0\) at \(x=0\), propagates through states that continue to advance in length (green nodes), and is absorbed once extrusion becomes fully blocked (gray nodes). The loop-length distribution \(N(x)\) is therefore the survival probability after \(m=x/\Delta x\) steps.

This mapping immediately explains the distribution classes: the number of propagating (green) nodes in \(\mathbf L\) equals the number of exponential modes in \(N(x)\). For one-sided extrusion (\(k=1\)) and $n$ obstacle classes, the \(\mathbf L\)-graph is star-like [Fig.~2(a), Fig.~S4], with a single propagating node \(A_{0}\). Once the process leaves this root node, loop-growth stops. The survival process therefore has a single mode, yielding the purely exponential law [Fig.~2(c)] for any $n$ and any parameters of the system, 
\begin{equation} 
N(x)=\lambda^{-1}e^{-x/\lambda}, \qquad (k=1). 
\label{eq:uni_exp} 
\end{equation}

For two-sided extrusion (\(k=2\)), by contrast, partially blocked states with one blocked and one mobile arm (\(A_b\), \(A_c\)) continue to propagate in length [Fig.~2(b)]. These extra propagating states generate extra decay modes. For \(n\) distinct barrier classes plus the extruder class, the distribution becomes a sum of \(n+2\) exponentials,
\begin{equation}
N(x)=\sum_{m=1}^{n+2} C_m e^{-\beta_m x/l_p},
\qquad (k=2),
\label{eq:multi_exp}
\end{equation}
with \(C_m>0\) and \(\beta_m>0\) fixed by the eigenvectors and eigenvalues of \(\mathbf L\).
Figure~2 shows the case of one infinitely long-lived barrier class
(\(n=1,\;\alpha=0\)) with varying density \(\gamma_1 \equiv \gamma_b\) and fixed cohesin
density \(\gamma=2.6\), corresponding to the experimental value
(Table~S2). Equation~\eqref{eq:multi_exp} then reduces to three distinct exponential modes depicted in Fig.~2(d). The analytical PDFs for both $k=1$ and $k=2$ agree well with direct stochastic simulations of active extrusion [Fig.~2(c,d), blue histograms].

Whereas the one-sided distribution is always monotonic and maximal at \(x=0\), Eq.~\eqref{eq:multi_exp} can develop a maximum at finite \(x>0\). Physically, this reflects the need for loops in two-sided extrusion to grow long enough to populate partially blocked states before final double arrest. At low obstacle density these states are rare, so \(N(x)\) smoothly reverts to the simple exponential form recovered in the obstacle-free limit. Peak formation for $k=2$ is controlled by a sharp density threshold: for
long-lived barriers (\(\alpha\to0\)), the minimal barrier density is the inverse
golden ratio, \(\gamma_b^{\min}=(\sqrt{5}-1)/2\approx0.62\) \cite{suppl}, whereas for short-lived barriers (\(\alpha\to\infty\)) it rises to \(\gamma_b^{\min}=1\).
Thus, once a cohesin is expected to encounter at least one static barrier during its lifetime, a peak is guaranteed, independent of \(\alpha\) and \(\gamma\). A peaked \(N(x)\) is therefore a generic consequence of sufficiently strong disorder in two-sided extrusion.

\begin{figure}[t]
    \centering  \includegraphics[width=0.5\textwidth]{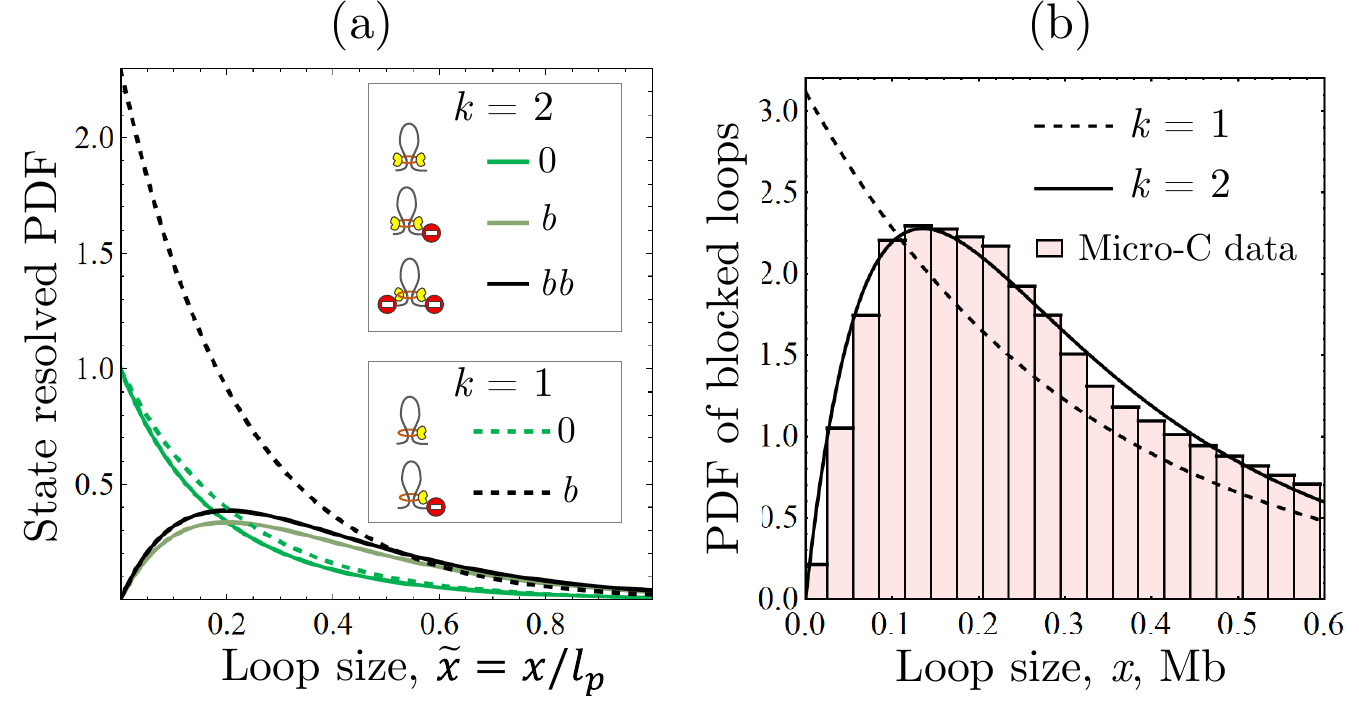}
    \caption{Blocked-loop statistics provide an experimental discriminator of extrusion symmetry.
    (a) Conditional loop-size distributions of individual states. For $k=2$ the free state $A_{0}$ decreases monotonically with $\tilde{x}=x/l_p$, whereas blocked and mixed states ($A_b$, $A_{bb}$) develop maxima at finite $\tilde{x}$. Dashed curves show the corresponding one-sided exponential laws ($k=1$).
    (b) Experimental distribution of CTCF--CTCF loop lengths (bars) from \cite{sept2025}, compared to the theoretical PDF $N_{\mathrm b}(x)$ of blocked loops with fitted parameters ($\gamma = 2.45$, $\gamma_{b} = 1$) whose values fall inside the experimental data range. The two-sided prediction (solid) reproduces the observed peak, whereas the one-sided prediction (dashed) remains exponential and fails to capture this essential feature.}

    \label{fig:fig3}
\end{figure}

The decisive observable, however, is not the full ensemble \(N(x)\), which is difficult to access experimentally, but the conditional distribution $N_{\mathrm b}(x)$ of blocked loops. This quantity provides a direct discriminator of extrusion symmetry. For a specific blocking protein, such as CTCF, we define
\begin{equation}
N_{\mathrm b}(x)\propto \sum_{q\in \mathrm{blocked}} A_q(x),
\qquad
\int_0^\infty N_{\mathrm b}(x)\,dx=1.
\end{equation}
For one-sided extrusion, this conditional distribution remains exponential, as in the full ensemble. For two-sided extrusion, by contrast, only the free state \(A_{0}\) decreases monotonically, whereas states containing at least one blocked arm develop maxima at finite \(x\) [Fig.~3(a)]. Clearly, this behavior is not restricted to strictly symmetric two-sided extrusion: alternative two-sided mechanisms, including stochastic arm switching or blockage-induced switching, are described by analogous length-state graphs \cite{suppl} and likewise generate peaked blocked-loop distributions. 
Physically, fully blocked states can be populated only after a finite growth stage; accordingly, \(N_{\mathrm b}(x)\) is generically peaked whenever both arms become active, even transiently.

Figure~3(b) compares the conditional blocked-loop PDF with the measured
distribution of CTCF--CTCF loop sizes from Micro-C/HiChIP data \cite{sept2025}. The
experimental histogram exhibits a pronounced peak, reproduced by the two-sided
prediction, whereas the one-sided \(N_{\mathrm b}(x)\) remains exponential and
fails to capture this qualitative feature. 
Blocked-loop statistics therefore
provide the central experimental discriminator of the theory: they rule out a
purely one-sided mechanism and require that both cohesin arms actively operate in vivo.

\textit{Synchronization of cohesin arms}. 
In the limit of dense, long-lived barriers (\(\gamma_b\to\infty,\;\alpha\to0\)), several decay modes in Eq.~\eqref{eq:multi_exp} become degenerate and \(N(x)\) approaches a \(\Gamma\)-distribution with shape parameter \(2\), as expected if the total loop length is the sum of two independent exponential arm extensions [Fig.~2(d)]. For any finite barrier density or persistence, however, the two arm extensions remain correlated. We therefore ask what controls this residual synchronization.

To quantify the internal dynamics of a two-sided extruder, we introduce the vector \(\mathbf{A}(r,\ell)\) of joint state-resolved densities for the right (\(r\)) and left (\(\ell\)) arm extensions. Its spatial moments define the Pearson coefficient, which we use as a measure of arm synchronization:
\begin{equation}
\rho=\frac{\langle \ell r\rangle-\langle \ell\rangle^2}{\langle \ell^2 \rangle-\langle \ell \rangle^2};
\qquad
\langle \ell\rangle=\langle r\rangle.
\end{equation}

A useful reference case is \(\gamma_b=0\), where the only obstacles are other cohesins. Even in the sterically crowded limit \(\gamma\to\infty\), the two arms do not become independent. Instead,
\begin{equation}
\rho_{\min}=\lim_{\gamma\to\infty}\rho(\gamma)=1/4,
\label{14}
\end{equation}
so cohesin crowding alone can desynchronize the arms only down to a finite floor \cite{suppl}.

The behavior becomes qualitatively richer in the presence of static barriers [Fig.~4]. For long-lived barriers (\(\alpha<1\)), increasing cohesin density can either decrease or increase \(\rho\), depending on the barrier density $\gamma_b$. The sign change occurs at
\begin{equation}
\gamma_b^{\mathrm{cr}}
=
\frac{3(1+\alpha)^2(1+2\alpha)}{1+2\alpha-3\alpha^2}.
\label{eq:gbcrit}
\end{equation}
For weak disorder, \(\gamma_b<\gamma_b^{\mathrm{cr}}\), increasing cohesin density decreases \(\rho\): the system is in a desynchronizing regime, where additional cohesins drive the arms toward the steric bound \(1/4\), Eq.~\eqref{14}. For strong disorder, \(\gamma_b>\gamma_b^{\mathrm{cr}}\), the trend reverses: increasing cohesin density raises \(\rho\), and the system enters a synchronizing regime in which long-lived pauses imposed by static barriers suppress correlations more strongly than cohesin crowding does. Only in this regime does the fully decorrelated \(\Gamma\)-limit become accessible, namely for \(\gamma_b\to\infty\) and \(\alpha\to0\).

The origin of this synchronization--desynchronization transition is simple. If static barriers alone reduce the arm correlation below the cohesin-controlled asymptote \(1/4\), then adding cohesins must increase \(\rho\), because long-lived roadblocks are statistically replaced by shorter-lived cohesin encounters ($\alpha<1$). If instead the barrier-only system is above \(1/4\), adding cohesins can only reduce \(\rho\).

The minimal transition density is reached for infinitely long-lived barriers, \(\alpha=0\), where \(\gamma_b^{\mathrm{cr}}=3\); for \(0<\alpha<1\), it increases continuously within the range $3<\gamma_b^{\mathrm{cr}}<\infty$.
Thus \(\gamma_b<3\) always places the system in the desynchronizing regime (weak disorder), irrespective of barrier persistence. For \(\alpha>1\), the transition disappears altogether: short-lived barriers cannot suppress arm--arm correlations below the critical bound. The experimentally relevant G1 regime lies in the weak-disorder, desynchronizing area [Fig.~4(a)]. CTCF barriers are relatively short-lived (\(\alpha_{\mathrm{CTCF}}>1\)) and too sparse (\(\gamma_{\mathrm{CTCF}}\approx1<3\)) to induce the transition. MCM complexes are much longer-lived (\(\alpha_{\mathrm{MCM}}\approx0\)), but their density is also subcritical (\(\gamma_{\mathrm{MCM}}\approx1.3<3\)).

\begin{figure}[t]
    \centering
  \includegraphics[width=0.48\textwidth]{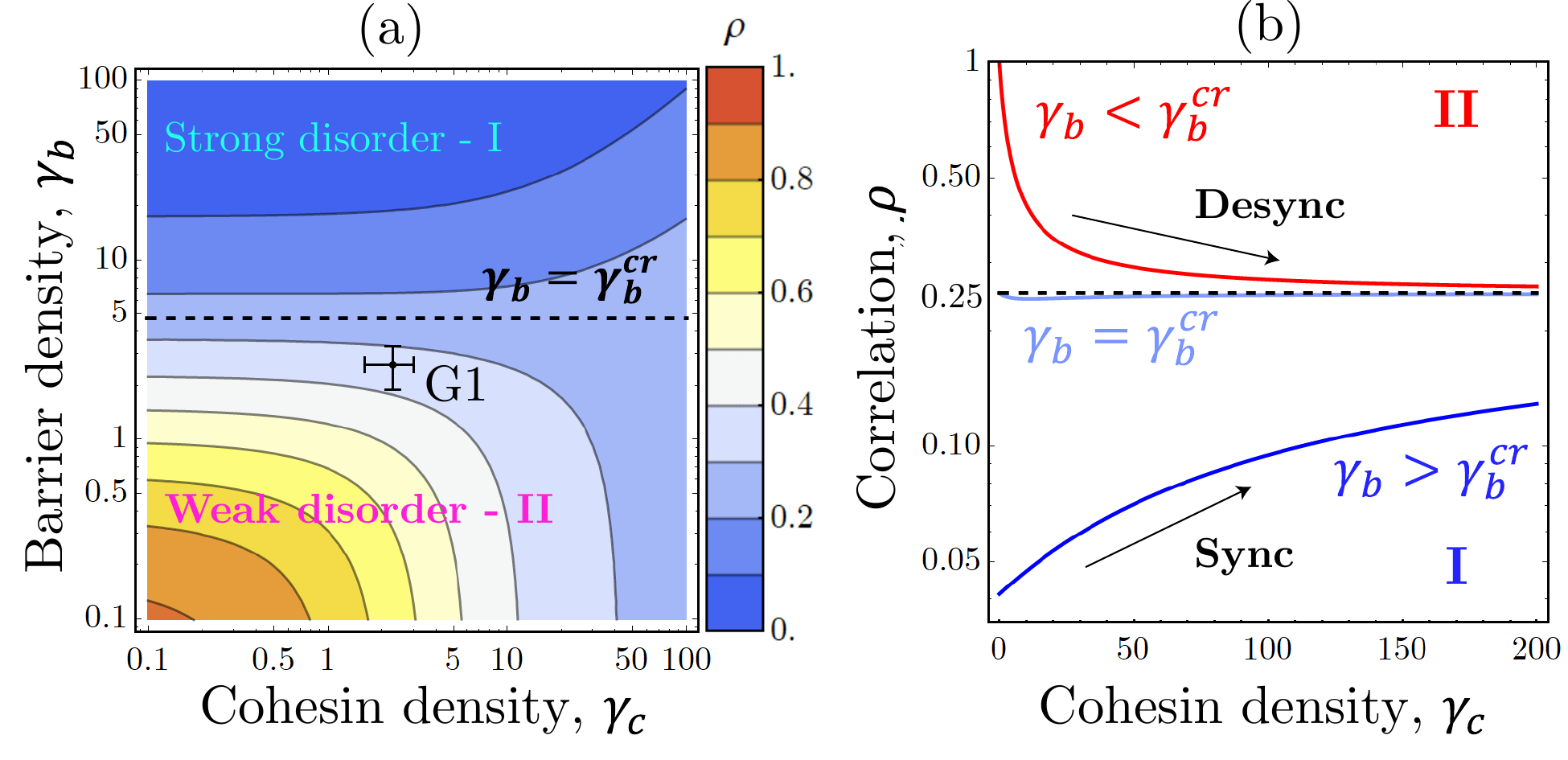}   
    \caption{Desynchronization regimes of cohesin arms in two-sided extrusion. 
    The two obstacle classes are other cohesins, with density \(\gamma_c=\gamma+\gamma_{\text{dyn}}\), and static barriers, with density \(\gamma_b\); barrier persistence is fixed at \(\alpha=0.2\).
    (a) Pearson correlation coefficient \(\rho\) between the left and right arm extensions in the \((\gamma_c,\gamma_b)\) plane. The dashed line marks the critical density \(\gamma_b^{\mathrm{cr}} \approx 4.7\) (see Eq.~\eqref{eq:gbcrit}) separating the strong-disorder regime (I) from the weak-disorder regime (II). The biologically relevant G1 point lies in regime II.
    (b) Correlation coefficient \(\rho\) versus cohesin density \(\gamma_c\) for representative barrier densities below (\(\gamma_b=0\), red), at (\(\gamma_b=4.7\), light blue), and above (\(\gamma_b=100\), blue) \(\gamma_b^{\mathrm{cr}}\). Accordingly, increasing cohesin density synchronizes the arms in regime I, leaves $\rho$ nearly unchanged at criticality, and desynchronizes them in regime II.}
        \label{fig:fig4}
\end{figure}

\textit{Conclusion}. 
We developed a mean-field theory of stationary loop statistics for active loop extrusion in the presence of kinetic barriers. The theory yields a universal expression for the mean loop size (Eq.~\eqref{eq:lambda_universal}), a graph-based classification of loop-length distributions (Fig.~2), and a kinetic description of a
disorder-induced synchronization--desynchronization transition between the two
extrusion arms (Fig.~4). In particular, the mean loop size is reduced relative to the bare processivity by a renormalized obstacle density \(\gamma_{\mathrm{eff}}\) that combines barrier abundance, lifetime, permeability, and extruder crowding. For one-sided extrusion (\(k=1\)), the loop-size distribution remains universally exponential, extending beyond the obstacle-free case. 

By contrast, two-sided extrusion (\(k=2\)) generically produces a multiexponential distribution, with the number of modes equal to the number of static barrier classes plus two (Eq.~\eqref{eq:multi_exp}, Fig.~2). Comparison with chromatin-contact data from \cite{sept2025} [Fig.~3(b)] shows that the observed peaked distribution of CTCF-anchored loops is incompatible with purely one-sided extrusion and therefore requires a mechanism in which both cohesin arms become active, at least transiently, in vivo.
This conclusion is consistent with recent observations of arm switching by cohesin \cite{Barth25}, and
parallels the general principle established for condensin-driven mitotic compaction: strictly one-sided extrusion leaves unavoidable gaps, and therefore cannot account for the \(\sim10^3\)-fold compaction of mammalian mitotic chromosomes \cite{Banigan19,banigan2020}. 

As a consequence, our analysis of arm--arm correlations shows that the response to changing extruder density \(\gamma\) depends qualitatively on the disorder regime, a feature that may become relevant across cell-cycle transitions and changing cellular states. Available data for CTCF \cite{brunner25,holzmann19,davidson2023} and MCM \cite{dequeker22,kulak2014minimal} suggest that interphase chromatin resides in the weak-disorder regime [Fig.~4(a)], where the arm--arm correlation has a strict lower bound of \(1/4\). This residual synchronization is expected to matter for loop-extrusion-mediated enhancer--promoter search \cite{ubertini25} and DNA repair \cite{arnould21,yang23}. 
More generally, the framework shows how the parameters of one-dimensional
kinetic disorder give rise to distinct statistical regimes, highlighting loop
extrusion as a versatile physical mechanism of genome organization and control.

\textit{Acknowledgements}. We gratefully acknowledge M. Kardar, L. Mirny and P. Virnau for illuminating discussions and A. Cherstvy and A. Goychuk for valuable comments on the manuscript. The work of A.C. and K.E.P. was supported by the Russian Science Foundation (Grant No. 25-13-00277). R.M. acknowledges support from DFG (CRC Data assimilation, Grant 318763901). K.E.P. also acknowledges support from the Alexander von Humboldt Foundation. 

\bibliography{shorttitles_updated,bibliography_converted}

\end{document}